\documentclass[runningheads]{llncs}
\usepackage{graphicx}
\usepackage{amsmath}
\usepackage{amssymb}
\usepackage{algorithm}
\usepackage{algorithmic}

\newcommand{\hnu}{\hat{\nu}}
\newcommand{\hv}{\hat{v}}
\renewcommand{\div}{\nabla\cdot\,}

\begin{document}
\title{Fisher-Rao Regularized Transport Analysis of the Glymphatic System and Waste Drainage}
\titlerunning{Fisher-Rao OMT and Glymphatic System}
%
\author{Rena Elkin\inst{1} \and Saad Nadeem\inst{1} \and Hedok Lee\inst{2} \and \\Helene Benveniste\inst{2} \and Allen Tannenbaum\inst{3}}
%

\authorrunning{R. Elkin \emph{et al.}}
%
\institute{Department of Medical Physics, Memorial Sloan Kettering Cancer Center, \\New York NY 10065, USA \and Department of Anesthesiology, Yale School of Medicine, \\New Haven, CT 06519, USA \and Departments of Computer Science and Applied Mathematics \& Statistics, \\Stony Brook University, Stony Brook, NY 11794, USA}
%
\maketitle              
\begin{abstract}
In this work, a unified representation of all the time-varying dynamics is accomplished with a Lagrangian framework for analyzing Fisher-Rao regularized dynamical optimal mass transport (OMT) derived flows. While formally equivalent to the Eulerian based Schr\"odinger bridge OMT regularization scheme, the Fisher-Rao approach allows a simple and interpretable methodology for studying the flows of interest in the present work. The advantage of the proposed Lagrangian technique is that the time-varying particle trajectories and attributes are  displayed in a single visualization. This provides a natural capability to identify and distinguish flows under different conditions. The Lagrangian analysis applied to the glymphatic system (brain waste removal pathway associated with Alzheimer's Disease) successfully captures known flows and distinguishes between flow patterns under two different anesthetics, providing deeper insights into altered states of waste drainage. 
\end{abstract}
\section{Introduction}
\label{sec:introduction}
\begin{figure}[t!]
 \centering
    \includegraphics[width=0.75\textwidth]{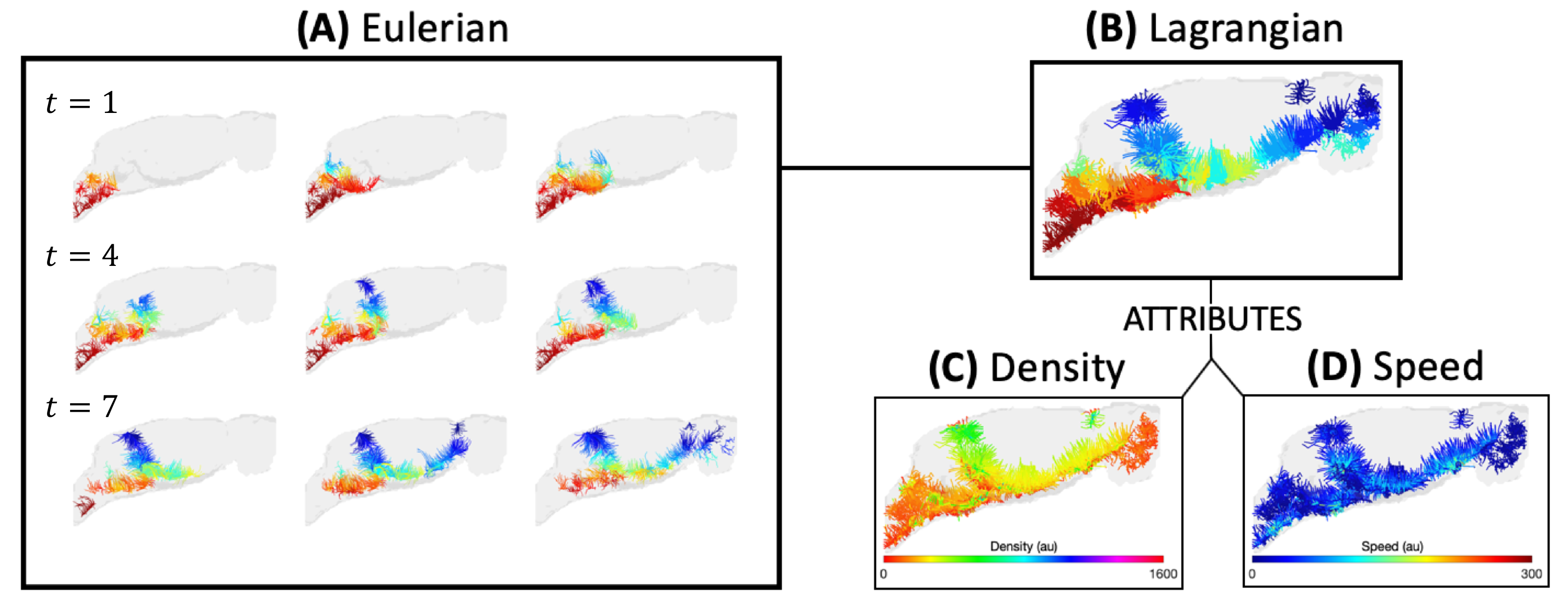}
    \caption{
    Eulerian and Lagrangian visualizations of flow dynamics. (A) For each time point, streamlines are computed at a fixed set of initial locations. Color is used to distinguish streamline clusters and 
    time-varying flow behavior is observed by comparing streamlines across time. 
    (B) Alternatively, temporal changes in particle trajectories are encompassed
    by pathlines 
    and presented in a single image. Time-varying particle attributes associated with the pathlines, such as (C) density and (D) speed, 
    exemplify
    the simplistic yet informative nature of our unified visualization framework.}
    \label{fig:EL}
\end{figure}

The optimal mass transport problem (OMT) entails minimizing the transportation cost associated with redistributing one probability distribution to match another. The static OMT formulation has been applied to medical image processing and registration \cite{feydy2017optimal,gerber2018exploratory,haker}.
The dynamical version of OMT \cite{French} opened up new possibilities for numerical solutions and extensions using tools from the field of fluid dynamics \cite{Buttazzo,papadakis2014optimal}.
In particular, the authors consider a regularization of the dynamical OMT formulation \cite{French} for modeling apparent fluid flow in dynamic contrast enhanced MRI (DCE-MRI) images where a diffusion term has been added to the Euler equation \cite{glymphatics,elkin2019optimal}. Contrast transport derived using earlier models without a diffusion term were vulnerable to noise in the data \cite{Ratner}, making the  inclusion of a diffusion term advantageous from both physical and numerical standpoints.
Analysis of the resulting velocity fields is typically carried out in an Eulerian framework, i.e., flow properties are considered at specific locations for each time point as compared to the Lagrangian framework, which tracks specific particles as they move over time; see Figure \ref{fig:EL}.

The Lagrangian approach gives a useful alternative to the Eulerian scheme and has recently been presented in conjunction with the aforementioned regularized version of OMT \cite{koundal2020optimal}. The pipeline taken from this work has untapped practical applications for distinguishing elusive physiologically relevant flow patterns and so we adopt it in the present work to study the glymphatic system (GS) and brain waste drainage. In \cite{koundal2020optimal}, the authors employ a regularized version of OMT by adding a diffusion term to the continuity equation in the Benamou-Brenier formulation of OMT \cite{French}. 
It turns out that there are two equivalent ways to derive the Lagrangian coordinates of the optimal trajectories. 
The first is via a transformation of the advection-diffusion equation in the regularized problem, 
and the second by adding a Fisher-Rao information type term to the kinetic energy in the original dynamic formulation \cite{Pavon}. In the present work, we adopt the second method, i.e., Fisher-Rao regularization.
We should note that employing Lagrangian coordinates derives improved visualizations of GS function in a time efficient display and also reveals new disparate dynamic waste drainage features in two different states of arousal that were previously difficult to discern.

Waste products are removed from the brain through the GS, 
a peri-vascular transit passage for cerebrospinal fluid (CSF) which facilitates mixing of CSF with interstitial fluid \cite{iliff2012paravascular}. 
Impaired waste clearance due to glymphatic dysfunction has been associated with vascular dementia, Alzheimer's Disease (AD) \cite{peng2016suppression} and sleep deprivation \cite{shokri2018beta}. 
The GS accelerates waste clearance from the brain during slow wave sleep and with certain hypnotics when compared to other anesthetics and wakefulness \cite{xie2013sleep}.
A recent study 
showed that unconsciousness induced with dexmedetomidine and low dose isoflurane (DEXM+ISO) that mimics natural sleep enhances GS function to a greater extent when compared to deep anesthesia induced with isoflurane (ISO) \cite{benveniste2017anesthesia}. These results suggest that hypnotic drugs that promote `natural' sleep might be superior to deeper states of sleep/anesthesia regarding preservation of sleep like GS function. Novel ways to improve or maintain GS function for general brain health is urgently needed and dependent on developing robust analysis tools to quantify its function.\\

\noindent
The main contributions of this work are as follows:
\begin{enumerate}
    \item We directly use the Fisher-Rao regularization of the energy functional in visualizing the flows,   highlighting the framework's natural treatment of small diffusion values
    needed for studying the GS.
    \item We successfully differentiate between GS flow patterns under two different anesthetics and provide additional new insights, in particular, to the relationship between GS flow and solute clearance via pathways outside of the brain.
\end{enumerate}

\section{Background}\label{section:background}

\subsection{Regularized OMT}
As alluded to above, the problem of optimal mass transport (OMT) is concerned with moving mass from one site to another so that minimal `work' is expended, and mass is preserved, 
see \cite{Villani,villani2008optimal} for a complete set of references.
For our purposes with respect to medical images, we are interested in the dynamical OMT formulation due to Benamou and Brenier \cite{French}. They proposed an alternative numerical solution by introducing time and solving a partial differential equation constrained space-time minimization problem. This formulation has the geometric interpretation of finding geodesics \cite{Otto} between the given densities $\mu_0$ and $\mu_1$ in the space of probability densities and the trajectory of the transport is explicitly factored into the cost.
Explicit representation of the density's evolution suggests exciting capabilities for improved image registration techniques to account for dynamical aspects of physiological processes with additional aptitude for analyzing interesting time-varying phenomena. This is particularly powerful for medical applications where a complete physical model is often impractical to implement.
We consider the following modified version of Benamou and Brenier's OMT formulation where a 
diffusion term is added in the continuity equation. 
The following \textbf{\emph{regularized OMT}} problem over the normalized time interval $t \in [0,1]$ is used to motivate the Lagrangian framework for flow representation \cite{koundal2020optimal} that will be detailed in the next section:
\begin{eqnarray}
\inf_{\mu,\nu} \int_0^1 \int_\mathcal{R} && \mu(t,x) \|\nu(t,x)\|^2 \, dx \, dt\label{omt1} \\
{\rm subject \, to \,} && \partial_t\mu + \nabla\cdot(\mu {\nu}) = \epsilon\Delta \mu,\label{eq:ade2}\\
&& \mu(0, \cdot) = \mu_0(\cdot), \quad \mu(1, \cdot) = \mu_1(\cdot), \; \epsilon >0 \label{eq:regec}
\end{eqnarray}
where $\mu = \mu(t,x)\in \mathbb{R}$ is the density interpolant between the given densities $\mu_0$ and $\mu_1$, assumed to have the same total mass
over a bounded, connected subspace $\mathcal{R}$ of $\mathbb{R}^d$, $\nu=\nu(t,x)\in\mathbb{R}^d$ is the velocity and $\epsilon$ is the diffusivity. It is interesting to note that this may also be regarded as a reformulation of the Schr\"{o}dinger bridge problem \cite{Pavon}.

\section{Lagrangian coordinates}\label{section:lag_framework}
Following \cite{koundal2020optimal},  the optimal trajectory in Lagrangian coordinates for the regularized case is defined as follows. Let $\mu_{min}$ and $\nu_{min}$ denote the minimum arguments of the action (\ref{omt1}) subject to (\ref{eq:ade2}).
Define the augmented velocity
\begin{equation}
    \hnu(t,x) = \nu(t,x) - \epsilon\nabla\log\mu(t,x).\label{eq:augv}
\end{equation}
Noticing that
\begin{equation}
\nabla\cdot(\mu\hnu) = \nabla\cdot\left[(\nu-\epsilon\nabla\log\mu)\mu\right]
=\nabla\cdot(\mu\nu)-\epsilon \Delta\mu,
\end{equation}
leads to the following  conservation form of the constraint \eqref{eq:ade2}:
\begin{equation} \label{eq:ade3}  \partial_t \mu + \nabla\cdot(\mu \hnu) = 0.\end{equation}
Analogous to \cite{French}, 
the Lagrangian coordinates of the flow $X = X(t,x)$ corresponding to the minimizing velocity $\nu_{min}$ is 
the solution of the differential equation
\begin{equation} \label{eq:lagrange}
X(0,x) = x, \quad \partial_t X = \hnu_{min}(t,X(t,x)),
\end{equation}
where according to \eqref{eq:augv},
 $   \hnu_{min} = \nu_{min} - \epsilon \nabla\log\mu_{min}.$

\subsection{Fisher-Rao regularization}

It is very important to note that the regularized OMT problem is equivalent to a Fisher-Rao information theoretic regularization of OMT, which is very closely connected to the Sinkhorn approach \cite{cuturi2013sinkhorn}, and is in fact a dynamic formulation of the Schr\"{o}dinger bridge problem \cite{Pavon}. For small $\epsilon$ in the advection-diffusion equation above, Sinkhorn will become unstable, which is why one needs to employ a different approach in the medical imaging realm. 
We will therefore derive an equivalent Lagrangian formulation of regularized OMT via the Fisher-Rao regularized functional.  We should note that there are a number of works demonstrating the well-known equivalence of Fisher-Rao and OMT joint interpolation functional and regularized OMT including \cite{Pavon,LYO,Chizat,Liero}. We briefly sketch the proof from these works to highlight some of the key steps of going from the regularized OMT problem to the Fisher-Rao regularized one. We refer the reader to the latter works for more details and insights into the methodology.

Following the notation of the previous section and references \cite{Pavon,LYO,Chizat,Liero}, we claim that the problem defined by (\ref{omt1}--\ref{eq:regec}) is equivalent to
\begin{eqnarray}
\inf_{\mu,\hnu} \int_0^1 \int_\mathcal{R} && \mu(t,x) (\|\hnu(t,x)\|^2  + \epsilon^2 \|\nabla \log \mu(t,x)\|^2) \, dx \, dt + 2 \epsilon H(\mu_1 \| \mu_0)\label{eq:schroed}\\ 
{\rm subject \, to: \,} && \partial_t \mu + \nabla\cdot(\mu {\hnu}) = 0, 
\quad\mu(0, \cdot) = \mu_0(\cdot), \quad \mu(1, \cdot) = \mu_1(\cdot)\label{eq:ade4}, 
\end{eqnarray}
for time-varying densities $\mu=\mu(t,x)\in\mathbb{R}$ and velocities $\nu=\nu(t,x)\in\mathbb{R}^d$
where
\begin{equation}
H(\mu_1\| \mu_0) := \int_\mathcal{R} \mu_1 (x)  \log \mu_1(x) - \mu_0(x) \log \mu_0(x) \, dx.\end{equation}
Clearly, $H(\mu_1 \| \mu_0)$ is a constant. Next, noting that
\begin{equation}
\hnu (t,x) + \epsilon \nabla \log \mu(t,x) = \nu(x,t) \end{equation}
and
\begin{equation}
\int_0^1 \!\!\!\int_\mathcal{R} \!\!\mu\| \hnu  + \epsilon \nabla \log \mu\|^2 dx dt
=\! \int_0^1 \!\!\!\int_\mathcal{R} \!\!\mu\| \hnu \|^2  +2 \epsilon \mu \langle \hnu, \nabla \log \mu \rangle
+\epsilon^2\mu \|\nabla \log \mu\|^2 dx dt,\label{eq:sb}
\end{equation}
we just need to show that
\begin{equation}
J := \int_0^1 \int_\mathcal{R} \langle \mu\hnu, \nabla \log \mu \rangle \, dx \, dt = H(\mu_1 \| \mu_0) \end{equation}
to verify our claim.
However, using integration by parts and the constraints~\eqref{eq:ade4} 
we see that
\begin{equation}
J = -\int_0^1 \int_\mathcal{R} \nabla \cdot (\mu \hnu) \log \mu \, dx \, dt
= -\int_0^1 \int_\mathcal{R} \mu \partial_t(\log \mu) \, dx \, dt + H(\mu_1 \| \mu_0). \end{equation}
Finally,
\begin{equation}
\int_0^1\! \int_\mathcal{R}\! \mu \partial_t(\log \mu) \, dx \, dt\! = \!\!\int_0^1 \!\int_\mathcal{R}\! \mu \frac{\partial_t\mu}{\mu} \, dx \, dt  = \!\!\int_\mathcal{R}\!\! (\mu_1(x)\! -\! \mu_0(x) ) \, dx = 0
\end{equation}
since the total masses of $\mu_1$ and $\mu_0$ are equal.

Letting $\hnu_{min}$ denote the optimal vector field for (\ref{eq:schroed}-\ref{eq:ade4}), the Lagrangian coordinate solution is then
exactly that given by \eqref{eq:lagrange} above.
As previously noted, there are very fast algorithms for solving such entropic regularized versions of OMT for sufficiently large $\epsilon$ \cite{cuturi2013sinkhorn}. Since we need to consider small  $\epsilon$ values, we directly solve (\ref{eq:schroed}-\ref{eq:ade4}) based on the numerical scheme given in \cite{steklova} where the 
initial and final densities $\mu_0, \mu_1$ are taken from the given DCE-MRIs.
\subsection{Pathlines}
In order to represent the characteristics of the flow over all time in one comprehensive figure, we use the Lagrangian framework to construct what are commonly known as {\em \textbf{pathlines}}. A pathline $X(t,x)$ with initial position $X(0,x) = x$ is given by \eqref{eq:lagrange} and traces the trajectory of an individual particle through-out the time interval. As detailed in \cite{koundal2020optimal}, additional information such as $\%$-signal from baseline is used to determine regions where flow is more likely to occur. Throughout this region of interest, $p$ seeding points, denoted $x_l$, are selected uniformly. Pathlines are then computed by integrating the augmented velocity \eqref{eq:augv} with initial positions $x_l$. The speed $\|\nu(t,X(t,x_l)\|$
of each particle is simultaneously computed along each pathline, referred to as the {\em \textbf{speed pathline}} and denoted $X_\nu = X_\nu(t,x_l)$, by (tri)linearly interpolating the derived velocity field. 

Next, in order to extract pathlines that are representative of the flow behavior in specific anatomical regions, we subsequently cluster the pathlines by proximity 
using the QuickBundles algorithm \cite{quickbundles}. 
All computations were implemented in MATLAB and 
the Lagrangian analysis took approximately 1 minute to run for each of the 8 datasets (5 DEXM+ISO, 3 ISO) of resolution $128 \times 128 \times 128$. 

\section{Discussion}\label{section:discussion}
While streamlines and pathlines are interchangeable for steady flows (i.e. time independent), the same cannot be said for unsteady flows (i.e. time dependent). A novel aspect of the  dynamical OMT formulation \cite{French} is its explicit description of the density's time-varying evolution, 
suggesting the need for Lagrangian analysis of this behavior. 
The Lagrangian approach provides an elegant framework to observe various attributes such as speed and density along the pathlines. Having a single representation for the history of a particle and its attributes across time lends 
itself to a natural means for differentiating between flows under different conditions. 
We show the capability of this framework as it pertains to the GS and waste drainage.

GS transport can be observed with DCE-MRI in combination with administration of para-magnetic contrast agents (e.g. DOTAREM) into the CSF \cite{lee2015effect}. However, supplementary analysis is required 
to extract and distinguish characteristics of the GS transport patterns and flow. Current techniques for quantifying GS transport include assessment of brain parenchymal solute uptake or clearance \cite{iliff2012paravascular}, kinetic analysis \cite{lee2015effect},
and k-means cluster analysis \cite{iliff2013brain,jiang2017impairment}.
Kinetic 
analysis strategies provide a static `snapshot' of GS transport over 2-3 hours and 
provide global influx and efflux parameters however, regional information at the voxel level cannot be accurately derived due to heterogeneous transport kinetics across brain regions. 

Here, we present a straightforward demonstration of the equivalence between the regularized dynamical OMT problem and the Fisher-Rao regularized variant in deriving the Lagrange coordinates associated with the flow. The Fisher-Rao perspective offers an accessible alternative methodology and deeper insight into the regularized OMT formulation as  previously applied \cite{koundal2020optimal}.
We apply the Lagrangian pipeline to measure GS function based on DCE-MRIs acquired in rats while under two different states of unconsciousness - `light' sleep/hypnosis with DEXM+ISO versus `deep' sleep/anesthesia with ISO based on data from \cite{benveniste2017anesthesia}. Specifically, we tested the Lagrangian OMT framework's ability to differentiate GS transport and solute drainage via pathways in the brain proper as well as along cranial efflux routes between these two different states of arousal.

\begin{figure}[t!]
\begin{center}
\includegraphics[width=0.6\textwidth]{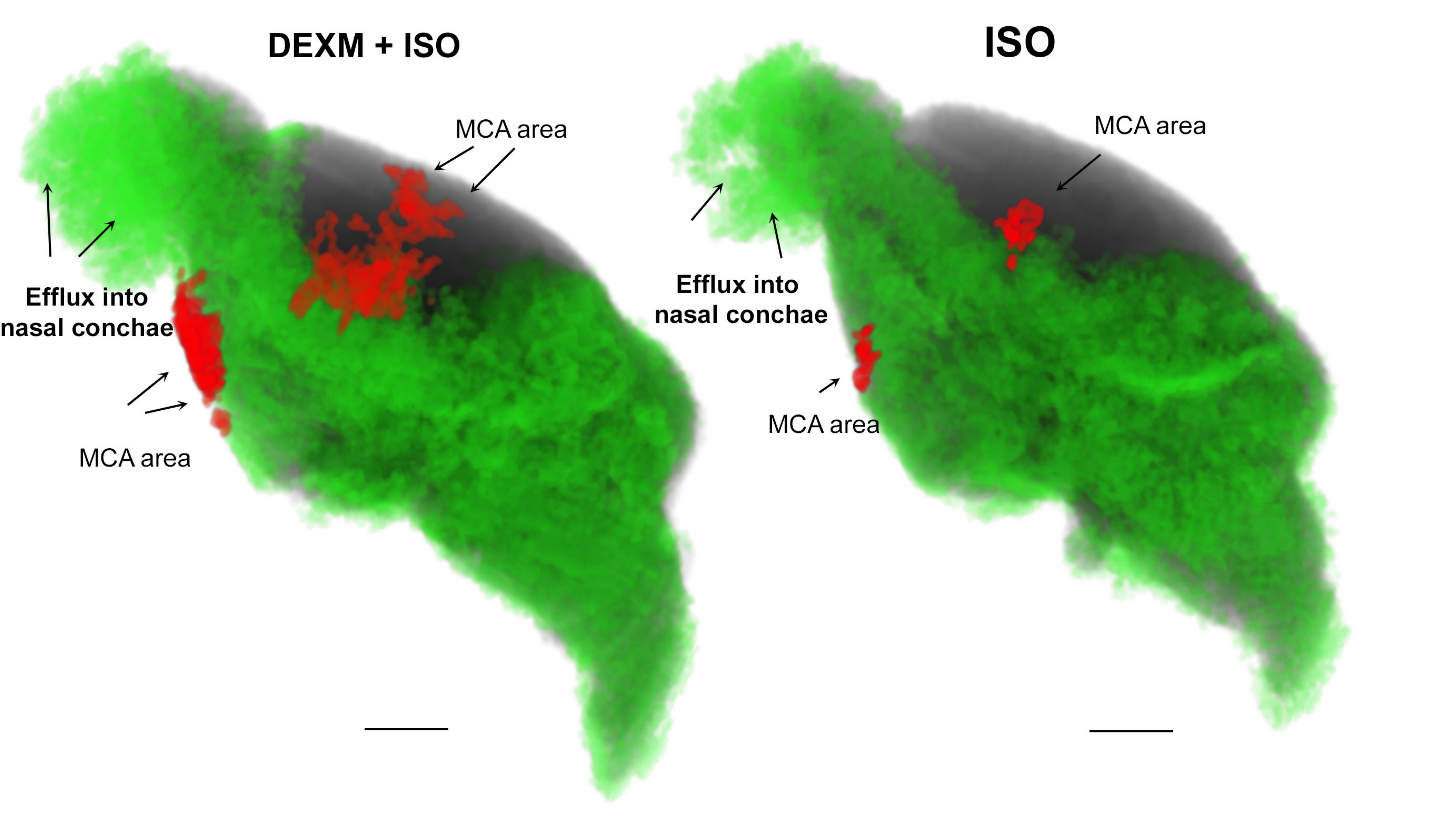}
\caption{Comprehensive pathline visualization of 130 minutes of glymphatic transport 
of DOTA (Gd-Dota) in a rat anesthetized with DEXM+ISO (left) and 
ISO (right). The pathlines in green are derived from the entire brain and nasal conchae. Both rats exhibit pathlines 
in the nasal conchae, 
signifying that Gd-DOTA exited along the olfactory nerves via the cribriform plate to lymphatic vessels in the submucosa of the nasal cavities. Pathlines that follow the middle cerebral artery (MCA) are outlined in red. More pathlines are clearly evident in the rat anesthetized with DEXM+ISO than ISO anesthesia, confirming that GS transport is more efficient during DEXM+ISO anesthesia. Scale bar = 3mm.}
\label{fig:1}
\end{center}
\end{figure}

Clustered pathlines derived from the Lagrangian framework are shown in Figure \ref{fig:1}. It is evident, based on the pathlines patterns observed in the rats under the two anesthetics, that GS transport function is different. With DEXM+ISO anesthesia, GS transport is accelerated (compared to ISO) as noted by the high density of pathlines extending into the parenchyma globally and along the middle cerebral artery (MCA). In contrast, the pathlines in rats anesthetized with ISO do not align with the MCA to the same extent 
and parenchymal uptake is also reduced in other areas including the midbrain and hippocampus. However, paradoxically, GS transport appears vigorous in the area of the olfactory bulb including solute efflux into the nasal conchae. 

\begin{figure}[t!]
\begin{center}
\includegraphics[width=\textwidth]{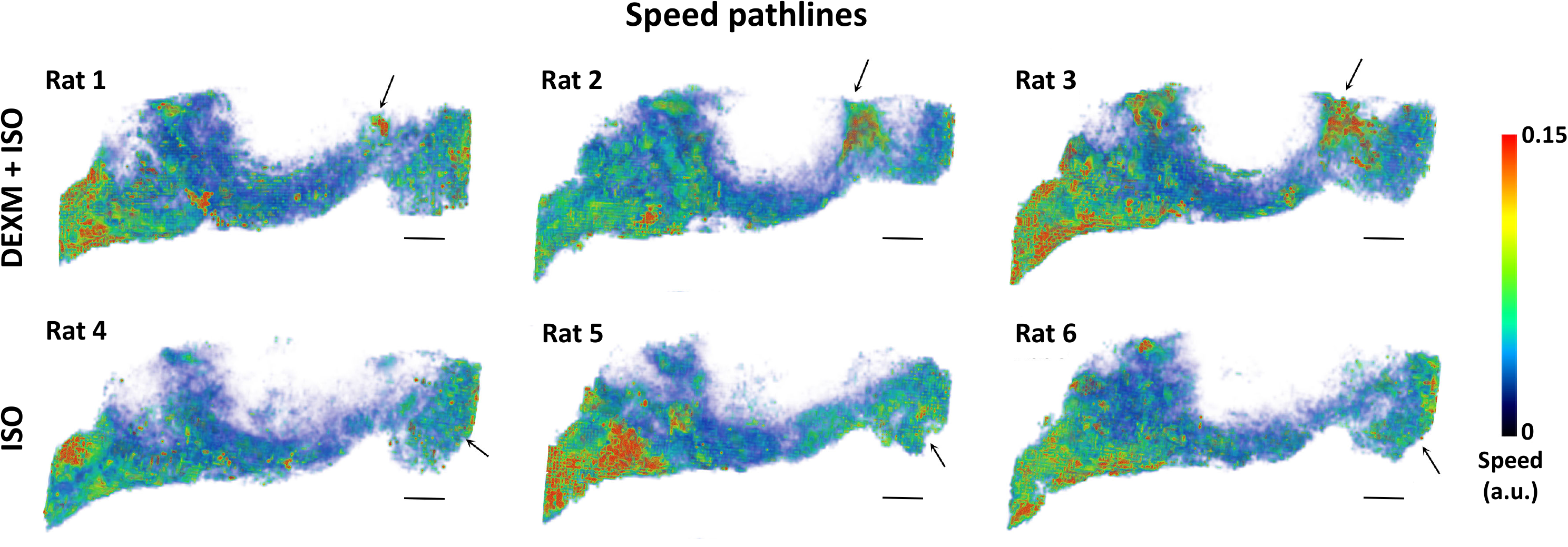}
\caption{Speed pathlines over 130 minutes of GS transport distinguishing flow behavior under two anesthetics, DEXM+ISO (top) and ISO (bottom). 
}
\label{fig:2a}
\end{center}
\end{figure}

\begin{figure}[t!]
\begin{center}
\includegraphics[width=\textwidth]{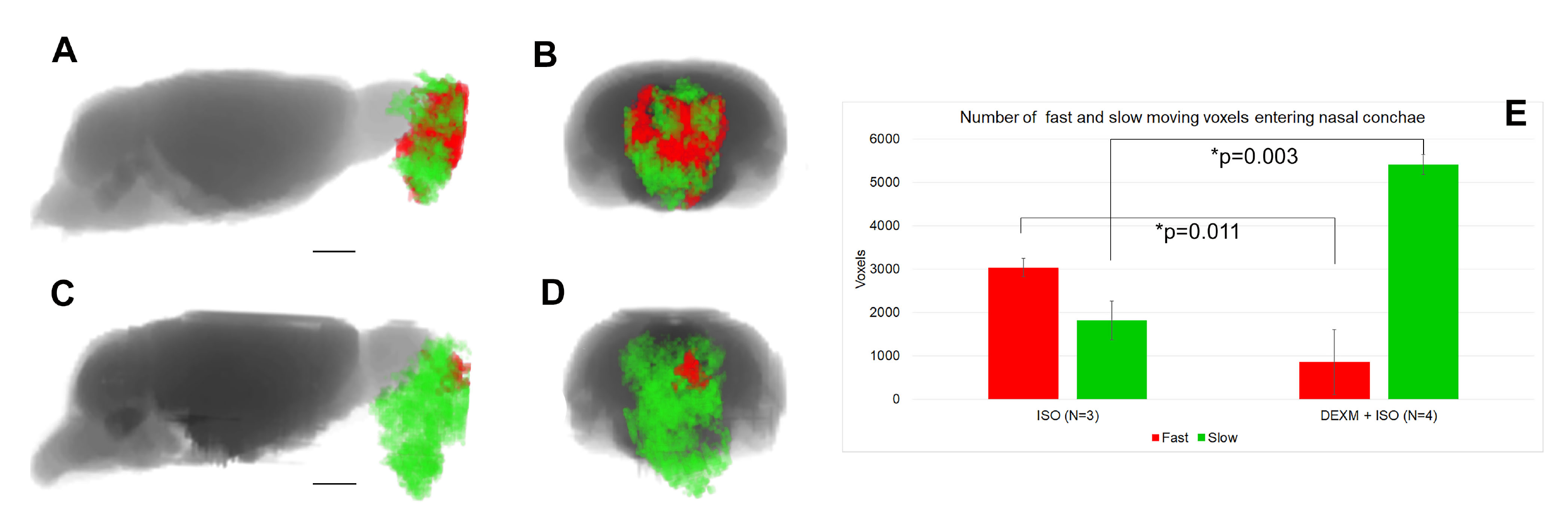}
\caption{Speed pathlines into the nasal conchae extracted from the GS speed `connectome' and divided up into fast (red) speed lines with speed magnitude $>0055$ a.u. and slower (green) speed lines with speed magnitudes in the range of $0.001-0.050$ a.u. (A, B) ISO anesthetized rats are characterized by rapid efflux of solute into the nasal cavity and (C, D) the vast majority of speed lines in the DEXM+ISO anesthetized rats are slower. 
(E) Proportion of voxels with fast moving solute. Data are mean $\pm$ SEM.}
\label{fig:2b}
\end{center}
\end{figure}

 Figure \ref{fig:2a} captures pathline speed (integrated over 130 minutes) through the GS in rats anesthetized with either DEXM+ISO or ISO. The speed pathlines associated with the whole brain are shown as a 3D volume rendered color-coded map. Higher and lower magnitudes of speed are represented by red and blue colors, respectively. More speed pathlines are apparent in the frontal part of the brain in the DEXM+ISO anesthetized rats compared to ISO. Paradoxically, in the ISO anesthetized state, pathline speed appears to be higher in the nasal conchae (arrows). In Figure \ref{fig:2b}, we have captured the faster (red) 
 and slower (green) moving speed lines, given in arbitrary units (a.u.) 
 in an ISO anesthetized rat (Figures \ref{fig:2b}A, B) compared to a DEXM+ISO anesthetized rat (Figures \ref{fig:2b}C, D). It is obvious that the proportion of faster moving particles in the clustered pathlines 
 is higher in ISO than DEXM+ISO anesthetized rats, as confirmed via quantitative analysis 
 (Figure \ref{fig:2b}E). 
 These differences suggest that solute efflux to lymphatic vessels in the nasal conchae may be acting as an efflux `valve' in the setting of impaired fluid flow in ISO anesthesia.

\section{Conclusion and Future Work}\label{section:future}
We presented a Lagrangian representation of Fisher-Rao regularized OMT to represent time-varying dynamics and various attributes in a single visualization. The resultant framework was then used to capture known GS transport flow patterns and provide additional insights in distinguishing GS function under different anesthetic conditions in addition to clearance of solutes via the olfactory nerves and into the nasal conchae. In the future, we will use the computed particle attributes as features 
in machine learning algorithms to differentiate and classify various transport related flow patterns for diagnostic purposes. 

\section*{Acknowledgements}
This study was supported by AFOSR grants (FA9550-17-1-0435, FA9550-20-1-0029), a grant from National Institutes
of Health (R01-AG048769, RF1-AG053991), MSK Cancer Center Support Grant/Core Grant (P30 CA008748), and a grant from Breast Cancer
Research Foundation (BCRF-17-193).

\bibliographystyle{splncs04}

\end{document}